\documentclass[prd,aps,twocolumn,preprintnumbers, showpacs, nofootinbib,superscriptaddress,notitlepage]{revtex4-1}
\usepackage{amsmath,graphicx,epsfig,amssymb}
\usepackage{inputenc}
\usepackage[usenames]{color}
\usepackage{ulem} 
\usepackage{bigstrut}
\usepackage{slashed}
\usepackage{multirow}
\usepackage{subfigure}
\usepackage{dsfont}
\usepackage{xcolor}

\allowdisplaybreaks

\newcommand{\nn}{\nonumber}

\begin{document}

\title{
Unpolarized Transverse-Momentum-Dependent Parton Distributions of the Nucleon from Lattice QCD}

\collaboration{\bf{Lattice Parton Collaboration ($\rm {\bf LPC}$)}}

\author{
Jin-Chen He}
\affiliation{Yang Yuanqing Scientiﬁc Computering Center, Tsung-Dao Lee Institute, Shanghai Jiao Tong University, Shanghai 200240, China}
\affiliation{Shanghai Key Laboratory for Particle Physics and Cosmology, Key Laboratory for Particle Astrophysics and Cosmology (MOE), School of Physics and Astronomy, Shanghai Jiao Tong University, Shanghai 200240, China}
\affiliation{Department of Physics, University of Maryland, College Park, MD 20742, USA}

\author{
Min-Huan Chu}
\affiliation{Yang Yuanqing Scientiﬁc Computering Center, Tsung-Dao Lee Institute, Shanghai Jiao Tong University, Shanghai 200240, China}
\affiliation{Shanghai Key Laboratory for Particle Physics and Cosmology, Key Laboratory for Particle Astrophysics and Cosmology (MOE), School of Physics and Astronomy, Shanghai Jiao Tong University, Shanghai 200240, China}

\author{Jun Hua} 
\affiliation{Guangdong Provincial Key Laboratory of Nuclear Science, Institute of Quantum Matter, South China Normal University, Guangzhou 510006, China}
\affiliation{Guangdong-Hong Kong Joint Laboratory of Quantum Matter, Southern Nuclear Science Computing Center, South China Normal University, Guangzhou 510006, China}

\author{Xiangdong Ji}
\affiliation{Department of Physics, University of Maryland, College Park, MD 20742, USA}

\author{Andreas  Sch\"afer}
\affiliation{Institut f\"ur Theoretische Physik, Universit\"at Regensburg, D-93040 Regensburg, Germany}

\author{Yushan Su}
\affiliation{Department of Physics, University of Maryland, College Park, MD 20742, USA}

\author{Wei Wang}
\email{Corresponding author: wei.wang@sjtu.edu.cn}
\affiliation{Shanghai Key Laboratory for Particle Physics and Cosmology, Key Laboratory for Particle Astrophysics and Cosmology (MOE), School of Physics and Astronomy, Shanghai Jiao Tong University, Shanghai 200240, China}
\affiliation{Southern Center for Nuclear-Science Theory (SCNT), Institute of Modern Physics, Chinese Academy of Sciences, Huizhou 516000, Guangdong Province, China}

\author{Yi-Bo Yang}
\affiliation{CAS Key Laboratory of Theoretical Physics, Institute of Theoretical Physics, Chinese Academy of Sciences, Beijing 100190, China}
\affiliation{School of Fundamental Physics and Mathematical Sciences, Hangzhou Institute for Advanced Study, UCAS, Hangzhou 310024, China}
\affiliation{International Centre for Theoretical Physics Asia-Pacific, Beijing/Hangzhou, China}
\affiliation{School of Physical Sciences, University of Chinese Academy of Sciences,
Beijing 100049, China}
 
\author{Jian-Hui Zhang}
\affiliation{School of Science and Engineering, The Chinese University of Hong Kong, Shenzhen 518172, China}
\affiliation{Center of Advanced Quantum Studies, Department of Physics, Beijing Normal University, Beijing 100875, China}

\author{Qi-An Zhang}
\email{Corresponding author: zhangqa@buaa.edu.cn}
\affiliation{School of Physics, Beihang University, Beijing 102206, China}

\begin{abstract}
We present a first lattice QCD calculation of the unpolarized nucleon's isovector transverse-momentum-dependent parton distribution functions (TMDPDFs), which are essential to predict observables of multi-scale, semi-inclusive processes in the standard model. We use a $N_f=2+1+1$ MILC ensemble with valence clover fermions on a highly improved staggered quark  (HISQ) sea to compute the quark momentum distributions in a large-momentum nucleon on the lattice. The state-of-the-art techniques in renormalization and extrapolation in the correlation distance on the lattice are adopted.  {The perturbative kernel up to next-to-next-to-leading order is taken into account}, and the dependence on the  pion mass and the hadron momentum is explored.   
Our results are qualitatively comparable with phenomenological TMDPDFs, which provide an opportunity to predict high energy scatterings from first principles. 
\end{abstract}

\maketitle

\section{ Introduction}

Since the nucleon is at the core of atoms and accounts for nearly all of the mass of the visible universe, exploring its internal structure has been a key task for more than a century in both particle and nuclear physics. In high-energy scattering, the quark and gluon transverse momentum and polarization degrees of freedom in the nucleon are best described by transverse-momentum parton distribution functions (TMDPDFs). Thus, mapping out the nucleon's TMDPDFs is a crucial step in understanding the interactions between quarks and gluons, and possibly the phenomenon of confinement~\cite{Collins:1981uk,Angeles-Martinez:2015sea}.  Moreover, predicting the observables in multi-scale, non-inclusive high energy processes such as semi-inclusive deep-inelastic scattering and Drell-Yan scattering at the large hadron collider (LHC) or electron ion collider (EIC) heavily relies on the knowledge of TMDPDFs~\cite{Collins:1984kg,Accardi:2012qut}. 

Whereas high energy experiments have accumulated a wealth of relevant data,  our knowledge of TMDPDFs is far from being complete. Their rapidity evolution, i.e. the Collins-Soper kernel~\cite{Collins:1981uk}, has been perturabtively calculated up to four loops~\cite{Moult:2022xzt,Duhr:2022yyp}, but TMDPDFs at low energies are nonperturbative in nature.  Based on thousands of data points from  the low-$p_T$  semi-inclusive DIS and Drell-Yan scattering processes and perturbative-QCD factorization,  a number of phenomenological analyses have been made to obtain state-of-art TMDPDFs~\cite{ Moos:2023yfa, Bury:2022czx, Bacchetta:2022awv, Scimemi:2019cmh, Bacchetta:2017gcc}. 
While similar datasets were employed in these analyses, the outcomes exhibit notable discrepancies. This suggests the presence of significant uncertainties in the global extraction of TMDPDFs, underscoring the need for additional constraints to achieve a more refined determination.

First-principles calculations of TMDPDFs require nonperturbative methods such as lattice QCD. A handful of available investigations using lattice QCD are limited to the ratios of moments of TMDPDFs \cite{Hagler:2009mb, Musch:2011er,Yoon:2015ocs,Yoon:2017qzo}. The development of large momentum effective theory (LaMET) allows the extraction of light-cone quantities through the simulation of equal-time quasi distributions~\cite{Ji:2013dva,Ji:2014gla}. A directly caluclation is TMDPDFs is non-trivial due to the presence of the soft function~\cite{Ji:2014hxa}, which involves two opposite light-like directions. Implementing this on an Euclidean lattice is a crucial difficulty.  Recent progress demonstrates that the rapidity-independent (intrinsic) soft function can be calculated from a large-momentum-transfer form factor of a light meson~\cite{Ji:2019sxk}, while the rapidity evolution kernel in the soft function can  be accessed via the quasi TMDPDFs/beam functions~\cite{Ji:2014hxa,Ebert:2019okf,Shanahan:2020zxr,Ji:2019ewn} or quasi transverse-momentum-dependent wave functions~\cite{Ji:2019sxk,Ji:2021znw}. 
Subsequent lattice efforts have been devoted to exploring the Collins-Soper kernel and intrinsic soft function. The agreement between lattice results and phenomenological analyses is encouraging~\cite{Shanahan:2020zxr,Schlemmer:2021aij,LatticeParton:2020uhz,Li:2021wvl,LPC:2022ibr}.

Following these developments, this work presents a first calculation of TMDPDFs from first principles. We simulate the TMD momentum distributions in a large momentum nucleon or quasi TMDPDFs on the lattice and perform a systematic study of renormalization properties by considering the subtractions from a combination of Wilson loop and short distance hadron matrix element~\cite{Zhang:2022xuw}. In the matching from quasi TMDPDFs, we include one-loop and two-loop perturbative contributions and employ the renormalization group equation to resum the logarithms. After analyzing the valence pion mass and momentum dependence, our final results for TMDPDFs are found to have a similar behavior as phenomenological fits.

The remainder of this paper is structured as follows.  Sec.~II presents the theoretical framework, followed by the presentation of Lattice simulations in Sec.~ III. Sec.~IV details the final results for TMDPDFs, while a concise summary and future prospects are outlined in Sec.~V.

\section{Theoretical framework}

\subsection{Constructing the equal-time quasi TMDPDFs}
    
Describing the momentum distributions of a parton inside a hadron, TMDPDFs $f(x, b_{\perp}, {\mu}, \zeta)$ are functions of the longitudinal momentum fraction $x$, the Fourier conjugate $b_{\perp}$ of the parton transverse momentum $q_{\perp}$, as well as the renormalization scale ${\mu}$ and the rapidity scale $\zeta$. In this work we will consider the flavor non-singlet/isovector unpolarized quark TMDPDFs, which do not mix with gluons. 

In LaMET, the correlations with modes traveling along the light-cone can be extracted from distributions in a fast-moving nucleon through large-momentum expansion. 
On the lattice, the equal-time quasi TMDPDFs are constructed as
\begin{align}
\tilde{f}_{\Gamma}&\left(x, b_{\perp}, P^z, {\mu}\right) \equiv \lim_{a\to0 \atop L\to\infty} \int\frac{dz}{2\pi} e^{-iz\left(xP^z\right)} \nn\\
& \times
\frac{\tilde{h}^{0}_{\Gamma}\left(z, b_{\perp}, P^z, a, L\right)}{\sqrt{Z_E\left(2L+z, b_{\perp}, a\right)} Z_O\left(1/a, {\mu}, \Gamma\right)},
\label{eq:momspacequasi}
\end{align}
where $a$ denotes the lattice spacing.  $\Gamma=\gamma^t$ or $\gamma^z$ is the Dirac matrix that can be projected onto $\gamma^+$ in the large momentum limit.  The $\tilde{h}^{0}_{\Gamma}\left(z, b_{\perp}, P^z, a, L\right)$ is built with a gauge-invariant nonlocal quark bilinear operator as
\begin{eqnarray}
\tilde{h}^{0}_{\Gamma}\left(z, b_{\perp}, P^z, a, L\right)=\langle P^z |\tilde{O}^{0}_{\Gamma,\sqsubset}(z,b_{\perp},P^z;  L)|P^z\rangle,  \label{eq:bareqtmdpdfme}\\
\tilde{O}^{0}_{\Gamma,\sqsubset}(z,b_{\perp},  L)\equiv 
	\bar{\psi}(b_{\perp}\hat{n}_{\perp})\Gamma U_{\sqsubset, L}\left( b_{\perp}\hat{n}_{\perp},  z\hat{n}_z\right)\psi(z\hat{n}_z).  \label{eq:staplewilsonline}
\end{eqnarray}
 In the above, $|P^z\rangle$ denotes the unpolarized nucleon state and $L$ denotes the ``infinity"  that the gauge link can reach. The staple-shaped Wilson link   is chosen as
\begin{align}
	&U_{\sqsubset, L}\left( b_{\perp}\hat{n}_{\perp},  z\hat{n}_z\right)\equiv U_{z}^{\dagger}\left(  (z+L)\hat{n}_z+b_{\perp}\hat{n}_{\perp},  b_{\perp}\hat{n}_{\perp}\right) \nn\\
	&\quad\times U_{{\perp}}\left( (z+L)\hat{n}_z+b_{\perp}\hat{n}_{\perp},  (z+L)\hat{n}_z \right) U_{z}\left( (z+L)\hat{n}_z, z\hat{n}_z\right), \label{eq:stapleshapedWilsonLink}
\end{align}
in which  $U_z$ is the path-ordered Euclidean gauge link along the $z$-direction, and $U_{\perp}$ is  along the transverse direction at the ``infinity" position $(z+L)$ on the finite lattice. Their explicit forms are:
\begin{align}
	&U_{z}(\xi^z_1 \hat{n}_z+\xi_{\perp} \hat{n}_\perp, \xi^z_2\hat{n}_z+\xi_{\perp} \hat{n}_\perp) \nn\\
	&\qquad =\mathcal{P}\exp\left[ -ig\int_{\xi^z_1}^{\xi^z_2 } d\lambda \hat{n}_z\cdot A\left(\lambda \hat{n}_z+\xi_{\perp} \hat{n}_\perp\right) \right], \\
	&U_{\perp}(\xi^z \hat{n}_z+\xi_{\perp 1} \hat{n}_\perp, \xi^z \hat{n}_z+\xi_{\perp 2} \hat{n}_\perp) \nn\\
	&\qquad = \mathcal{P}\exp\left[ -ig\int_{\xi_{\perp 1}}^{\xi_{\perp 2}} d\lambda \hat{n}_\perp \cdot A\left( \xi^z \hat{n}_z + \lambda \hat{n}_{\perp} \right) \right]. 
\end{align}

\begin{figure}[!th]
\centering
\includegraphics[scale=0.45]{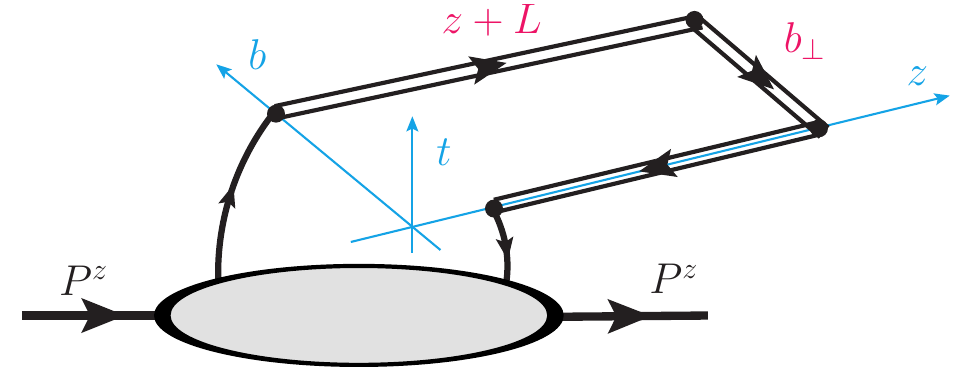}
\caption{Illustration of the  quasi TMDPDFs. }
\label{fig:quasiTMDPDF}
\end{figure}

An illustration of the quasi TMDPDFs is given in Fig.~\ref{fig:quasiTMDPDF}, in which
The staple-shaped Wilson link is depicted as double lines. 

\subsection{Renormalization of the TMD matrix elements}  \label{sec:theo_renormalization}

Quantities in Eq.~(\ref{eq:bareqtmdpdfme}) and (\ref{eq:staplewilsonline}) with the superscript ``$0$" are bare quantities on a finite lattice. They contain  linear divergence, pinch-pole singularity and logarithmic divergence. Both the linear divergence and pinch-pole singularity can be renormalized by the square root of  Wilson loop $\sqrt{Z_E(2L+z,b_\perp,a)}$ \cite{Ji:2017oey, Ishikawa:2017faj, Green:2017xeu, Ji:2020brr,Shanahan:2019zcq}; and the logarithmic divergence can be renormalized by {a factor $Z_O(1/a,\mu)$, which is extracted from matching between the lattice and perturbative calculation of zero-momentum matrix elements in the perturbative region} \cite{Ji:1991pr, LatticePartonCollaborationLPC:2021xdx,Ji:2021uvr,Zhang:2022xuw}.

The Wilson loop $Z_E(r=2L+z,b_{\perp},a)$ is defined as the vacuum expectation of a rectangular shaped space-like gauge links with size $r\times b_{\perp}$, {which can be written as
\begin{align}
    Z_E(r,b_{\perp},a) =& \left\langle U_{z}^{\dagger}\left(  (z+L)\hat{n}_z+b_{\perp}\hat{n}_{\perp},  (-z)\hat{n}_z+b_{\perp}\hat{n}_{\perp}\right) \right. \nn\\
    &\quad  \times U_{{\perp}}\left( (z+L)\hat{n}_z+b\hat{n}_{\perp},  (z+L)\hat{n}_z \right) \nn \\ 
    &\quad  \times  U_{z}\left( (z+L)\hat{n}_z, (-z)\hat{n}_z\right) \nn\\
    &\quad  \times \left. U_{{\perp}}^\dagger\left( (-z)\hat{n}_z+b_{\perp}\hat{n}_{\perp},  (-z)\hat{n}_z \right) \right\rangle. \label{eq:Wilson_Loop}
\end{align}}

The Wilson loop is introduced to eliminate the linear divergence of the form $e^{-\delta\bar{m}r}$, which comes from the self-energy corrections to the gauge link \cite{Ji:2017oey,LatticePartonCollaborationLPC:2021xdx}, as well as the pinch-pole singularity, which is related to the heavy quark effective potential term $e^{-V(b_{\perp})L}$ describing the interactions between the two Wilson lines along the z direction in the staple link \cite{Ji:2019ewn}.

The logarithmic divergence factor $Z_O$ can be extracted from the zero-momentum bare matrix elements $\tilde{h}_{\Gamma}^{0}\left(z, b_{\perp}, 0, a, L\right)$. In order to keep the renormalized matrix elements consistent  with  perturbation theory, $Z_O$ should be determined from the condition:
\begin{align}
    Z_O\left(\frac{1}{a}, \mu, \Gamma\right) =\lim _{L \rightarrow \infty} \frac{\tilde{h}_{\Gamma}^{0}\left(z, b_{\perp}, 0, a, L\right)}{\sqrt{Z_E\left(2L+z, b_{\perp}, a\right)} \tilde{h}_{\Gamma}^{\mathrm{\overline{MS}}}\left(z, b_{\perp}, \mu\right)}\label{eq:zo_definition}
\end{align}
in a specific window where $z \ll \Lambda_{\text{QCD}}^{-1}$ so that perturbation theory is valid. 
{The perturbative result  for the $\Gamma(=\gamma^t$ or $\gamma^z)$ zero-momentum matrix element up to one-loop order in the $\overline{\text{MS}}$ scheme reads }
\begin{align}
&\tilde{h}_{\Gamma}^{\overline{\text{MS}}}(z,b_{\perp}, \mu)=1+\frac{\alpha_s(\mu)C_F}{2\pi} \left[ \frac{1}{2}+ \right. \nn\\
   &\qquad \left.  \frac{3}{2}\ln\left( \frac{\mu^2(b_{\perp}^2+z^2)e^{\gamma_E}}{4} -2\frac{z}{b_{\perp}}\arctan\frac{z}{b_{\perp}}   \right)  \right].
\end{align}
Here the perturbation results have been evolved from the intrinsic physical scale $\mu_0=2 e^{-\gamma_{E}}/\sqrt{z^2+b_{\perp}^2}$ to the $\overline{\text{MS}}$ scale $\mu$ using the renormalization group equation~\cite{Su:2022fiu}. 

\subsection{Matching at next-to-next-to-leading order and renormalization group resummation}

It has been shown that quasi TMDPDFs have the same collinear degrees of freedoms as TMDPDFs~\cite{Ji:2019ewn}. Their differences from soft modes can be attributed to the intrinsic soft function and different rapidity scales. Also contributions from highly off-shell modes are local~\cite{Ebert:2019okf}. Thus TMDPDFs $f\left(x, b_{\perp}, \mu, \zeta\right) $ are connected to quasi TMDPDFs $\tilde{f}_\Gamma\left(x, b_{\perp}, \zeta_z, \mu\right)$ via a multiplicative  factorization~\cite{Ji:2019ewn,Ebert:2022fmh}: 
\begin{align}
\tilde{f}_\Gamma&\left(x, b_{\perp}, \zeta_z, \mu\right) \sqrt{S_I\left(b_{\perp}, \mu\right)}= H\left(\frac{\zeta_z}{\mu^2}\right) e^{\frac{1}{2} \ln \left(\frac{\zeta z}{\zeta}\right) K\left(b_{\perp}, \mu\right)} \nonumber\\
&\quad \quad\times f\left(x, b_{\perp}, \mu, \zeta\right)  
+\mathcal{O}\big(\frac{\Lambda_{\mathrm{QCD}}^2}{\zeta_z}, \frac{M^2}{\left(P^z\right)^2}, \frac{1}{b_{\perp}^2 \zeta_z}\big), \label{eq:factoizationFormula}
\end{align}
where $S_I$ denotes  the intrinsic soft function {that has been calculated on the lattice in Refs.~\cite{LatticeParton:2020uhz,Li:2021wvl,Chu:2023jia,LatticePartonLPC:2023pdv} }  and $K$ denotes the  Collins-Soper kernel.
Power corrections are suppressed by $\mathcal{O}\left({\Lambda_{\mathrm{QCD}}^2}/{\zeta_z}, {M^2}/{\left(P^z\right)^2}, {1}/(b_{\perp}^2 \zeta_z)\right)$, which implies that TMDPDFs can only be accurately obtained for moderate values of $x$.

The matching kernel $H$,  as a function of $\zeta_z/\mu^2=\left(2xP^z\right)^2/\mu^2$, has been perturbatively determined up to next-to-leading order (NLO) \cite{Ji:2019ewn,Ji:2018hvs,Ebert:2019okf, Ji:2020ect}
\begin{align}
H^{(1)}\left(\frac{\zeta_z}{\mu^2}\right)=\frac{\alpha_s C_F}{2 \pi}\left(-2+\frac{\pi^2}{12}+\ln \frac{\zeta_z}{\mu^2}-\frac{1}{2} \ln ^2 \frac{\zeta_z}{\mu^2}\right),
\end{align}
as well as next-to-next-to-leading order (NNLO), calculated recently \cite{delRio:2023pse, Ji:2023pba} 
\begin{align}
	H^{(2)}&\left(\frac{\zeta_z}{\mu^2}\right)=\alpha_s^2\left[c_2-\frac{1}{2}\left(\gamma_C^{(2)}-\beta_0 c_1\right) \ln \frac{\zeta_z}{\mu^2} \right. \nn\\
	& \quad \left.-\frac{1}{4}\left(\Gamma_{\text {cusp }}^{(2)}-\frac{\beta_0 C_F}{2 \pi}\right) \ln ^2 \frac{\zeta_z}{\mu^2}-\frac{\beta_0 C_F}{24 \pi} \ln ^3 \frac{\zeta_z}{\mu^2}\right],
\end{align}
where $\zeta_z=\left(2xP^z\right)^2$ and $c_2 = 0.0725 C_F^2 - 0.0840 C_F C_A + 0.1453 C_F n_f / 2$. The perturbative TMDPDF is calculated   in the $\overline{\mathrm{MS}}$ scheme with a fixed renormalization scale $\mu$, while the quasi TMDPDFs are associated with the Collins-Soper scale $\sqrt{\zeta_z}$, which is the intrinsic physical scale of  perturbative matching. In order to expose the intrinsic physical scale~\cite{Su:2022fiu}, we resum the large logarithms {$\sim\ln^n\left(\zeta_z/\mu^2\right)$ in the small $x$ region} through the renormalization group (RG) equation for $H$:
\begin{align}
    \mu^2 \frac{d}{d \mu^2} \ln H\left(\frac{\zeta_z}{\mu^2}\right)=\frac{1}{2} \Gamma_{\text {cusp }}\left(\alpha_s\right) \ln \frac{\zeta_z}{\mu^2}+\frac{\gamma_C\left(\alpha_s\right)}{2},
\end{align}
where $\gamma_C=2\gamma_F+\Gamma_S+2\gamma_H$ with $\gamma_C^{(1)}=-C_F/\pi$ and $\gamma_C^{(2)}=(a_1C_FC_A+a_2C_F^2+a_3C_Fn_f)$, the coefficients $a_1=44\zeta_3-\frac{11\pi^2}{3}- \frac{1108}{27}$, $a_2=-48\zeta_3 + \frac{28\pi^2}{3} - 8$ and $a_3=\frac{2\pi^2}{3}+\frac{160}{27}$ \cite{Ji:2019ewn,Ji:2020ect}. The cusp anomalous dimension $\Gamma_{\text {cusp }}$ is known up to the four-loop level for the quark case \cite{Moch:2017uml,Lee:2019zop,delRio:2023pse}.

\begin{figure}
\centering
\includegraphics[scale=0.8]{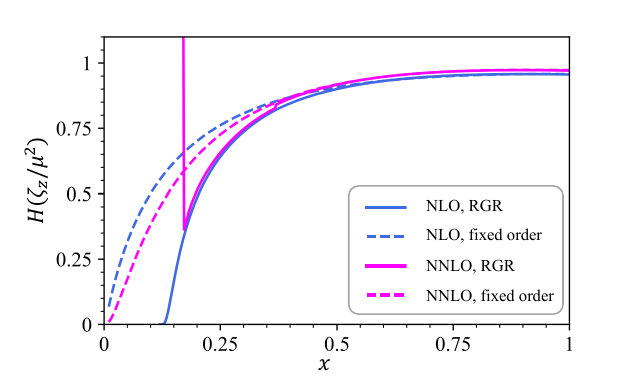}
\caption{Comparison of NLO and NNLO matching kernels at fixed order with $\mu=2$GeV (dashed lines) and running from Collins-Soper scale $\mu_0=2xP^z$ to $\overline{\mathrm{MS}}$ scale $\mu=2$GeV (solid lines), $P^z$ is chosen as $2.15$GeV.  The abrupt surge in behavior observed in NNLO+RGR is attributed to the presence of the Landau pole. }
\label{fig:Fig_matchingKernel_2}
\end{figure}

In practice, we compare the matching kernel at fixed order and with employing the RG resummation starting from the Collins-Soper scale $\mu_0=2xP^z$ to $\mu=2$GeV. After the resummation, the intrinsic scale $2xP^z$ appears in the running coupling $\alpha_s(2 x P^z)$.  
Fig.~\ref{fig:Fig_matchingKernel_2} shows the comparison of NLO and NNLO matching kernels with and without RG evolution. One can see that the RG evolution changes the perturbative behavior at small-$x$, and makes the predictions of the TMDPDFs in this region less reliable.

\section{Lattice simulations}

We use the valence tadpole improved clover fermion on the hypercubic (HYP) smeared ~\cite{Hasenfratz:2001hp} $2+1+1$ flavors MILC configurations with highly improved staggered quark (HISQ) sea and 1-loop Symanzik improved gauge action~\cite{MILC:2012znn}. We analyze a single ensemble with lattice spacing $a=0.12$~fm and volume $n_s^3\times n_t=48^3\times64$ using physical sea quark masses, and two choices of light valence quark mass corresponding to $m_{\pi}^{\mathrm{val}}=\{220,~310\}$ MeV. The HYP smearing is also used for nonlocal correlation functions 
to improve the statistical signal.
In order to explore  the momentum dependence, we employ three different nucleon momenta as $P^z=2\pi/(n_sa)\times\{8, 10, 12\}=\{1.72, 2.15, 2.58\}$~GeV.

We adopt momentum-smearing point source~\cite{Bali:2016lva}  at several time slices, and average  correlation functions for both the forward and backward directions in $z$  and transverse  space of  the gauge link. In total, there are  $1000$ (configurations) $\times 16$  (source time slices) $\times 4$ (forward/backward directions of the $z$ and transverse axes) measurements for the $m_{\pi}^{\mathrm{val}}=220$ MeV case and $1000\times 4\times 4$ measurements for the  $310$ MeV case.

\subsection{Dispersion relations}


\begin{figure*}[!th]
    \centering
    \includegraphics[width=1.0\textwidth]{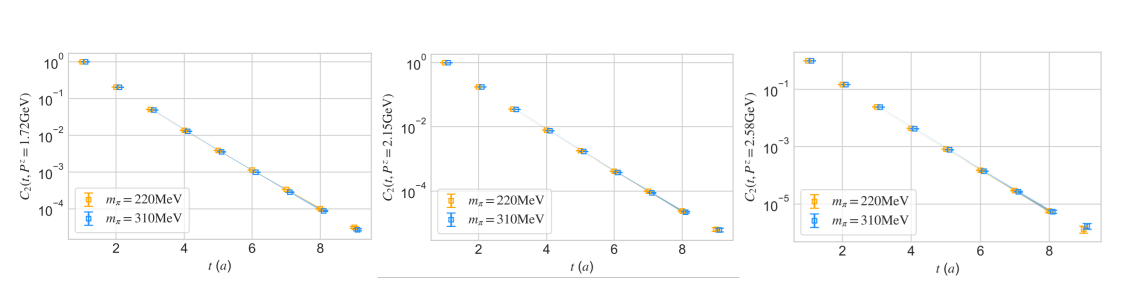} 
   \caption{Two-point correlation functions of nucleons with pion mass $m_{\pi}=220$~MeV and $310$~MeV at largest three momenta $P^z=1.72$GeV (left panel), $2.15$GeV (central panel) and $2.58$GeV (right panel). The colored bands indicate the fit ranges ($t\in[3,8]a$) and results of each two-state fits. }
   \label{fig:disp_fit2pt}
\end{figure*}

\begin{figure}[!th]
    \centering
    \includegraphics[width=0.4\textwidth]{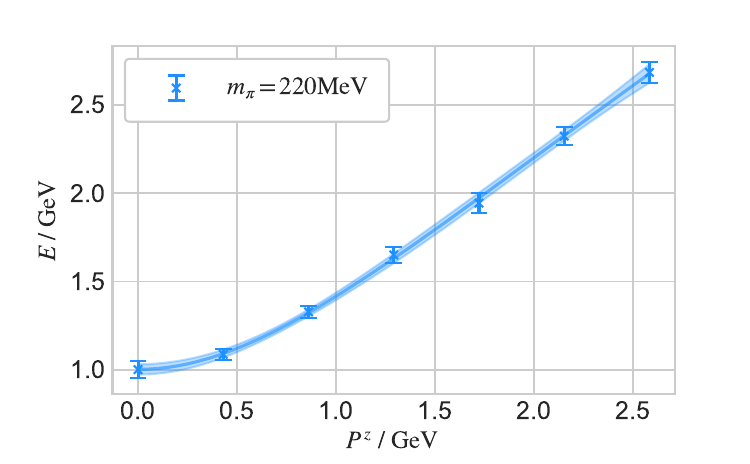} 
    \includegraphics[width=0.4\textwidth]{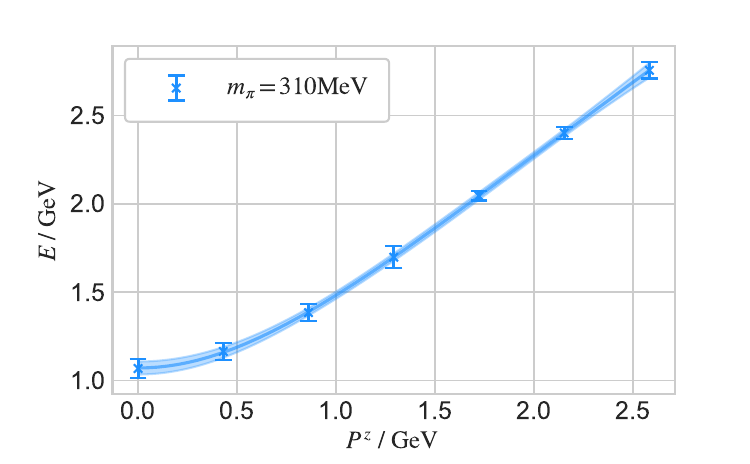}
    \caption{{The dispersion relations of nucleons with pion mass $m_{\pi}=220$~MeV (left panel) and $310$~MeV (right panel). The data with momentum up to $2.58$~GeV can be described by Eq.\ref{eq:dispersionrelation}. Both fit bands are consistent with the data points. }
   }
    \label{fig:disp}
\end{figure}

The two-point correlation functions of the nucleon are defined as
\begin{align}
&C_2(t, \vec{P}) = \Big\langle  \sum_{\vec{y}} e^{i\vec{P}\cdot\vec{y}} T\chi(\vec{y},t)\bar{\chi} (\vec{0},0)    \Big\rangle, \label{eq:2pt}
\end{align}
where $T=\left(1+\gamma^t\right)/2$ denotes the unpolarized projector, and $\chi=\epsilon^{abc}u_a \left(u_b^T C\gamma_5 d_c\right)$ is the nucleon interpolation field. We have computed the two-point functions using nucleon momenta up to $2.58$ GeV to examine the dispersion relation.

For the three momenta $P^z=1.72, 2.15, 2.52$ GeV, the statistics of the two-point correlation functions are 1000 (configurations) $\times16$ (source time slices) for $m_{\pi}=220$~MeV and $1000\times4$ for $m_{\pi}=310$~MeV; while for the cases with a momentum smaller than $1.72$~GeV, there are $1000\times4$ measurements for $m_{\pi}=220$~MeV and $1000\times1$ measurements  for $m_{\pi}=310$~MeV. Throughout we use the parametrization $C_2(t)=c_0e^{-E_0t}\left(1+c_1e^{-\Delta Et} \right)$, and we perform two-state fits to extract the ground-state energies, as shown in Fig.\ref{fig:disp_fit2pt}.

Based on the ground-state energies with different $P^z$, one can obtain the dispersion relations of the nucleons with the pion mass $m_{\pi}=220$~MeV and $310$~MeV. We adopt the following parametrization 
\begin{align}
E\left(P^z\right)=\sqrt{m^2+b_1\left(P^z\right)^2+b_2\left(P^z\right)^4a^2}, \label{eq:dispersionrelation}
\end{align}
where the quadratic term of lattice spacing $a$ is introduced to parameterize discretization errors.
The fit results are shown in  Fig.\ref{fig:disp}. 
For the $m_{\pi}=220$~MeV case, it is found that $b_1=1.014 (95) $ and $b_2=-0.014 (17)$, while for the  $310$~MeV case, the fit gives $b_1=1.066 (80)$ and $b_2=-0.015 (14)$.  From these results,  we can see that the dispersion relation is consistent with $E\left(P^z\right)=\sqrt{m^2+(P^z)^2}$ within uncertainties.

\subsection{Bare quasi TMDPDFs from correlated joint fits}

\begin{figure*}[!t]
\centering
\includegraphics[scale=0.72]{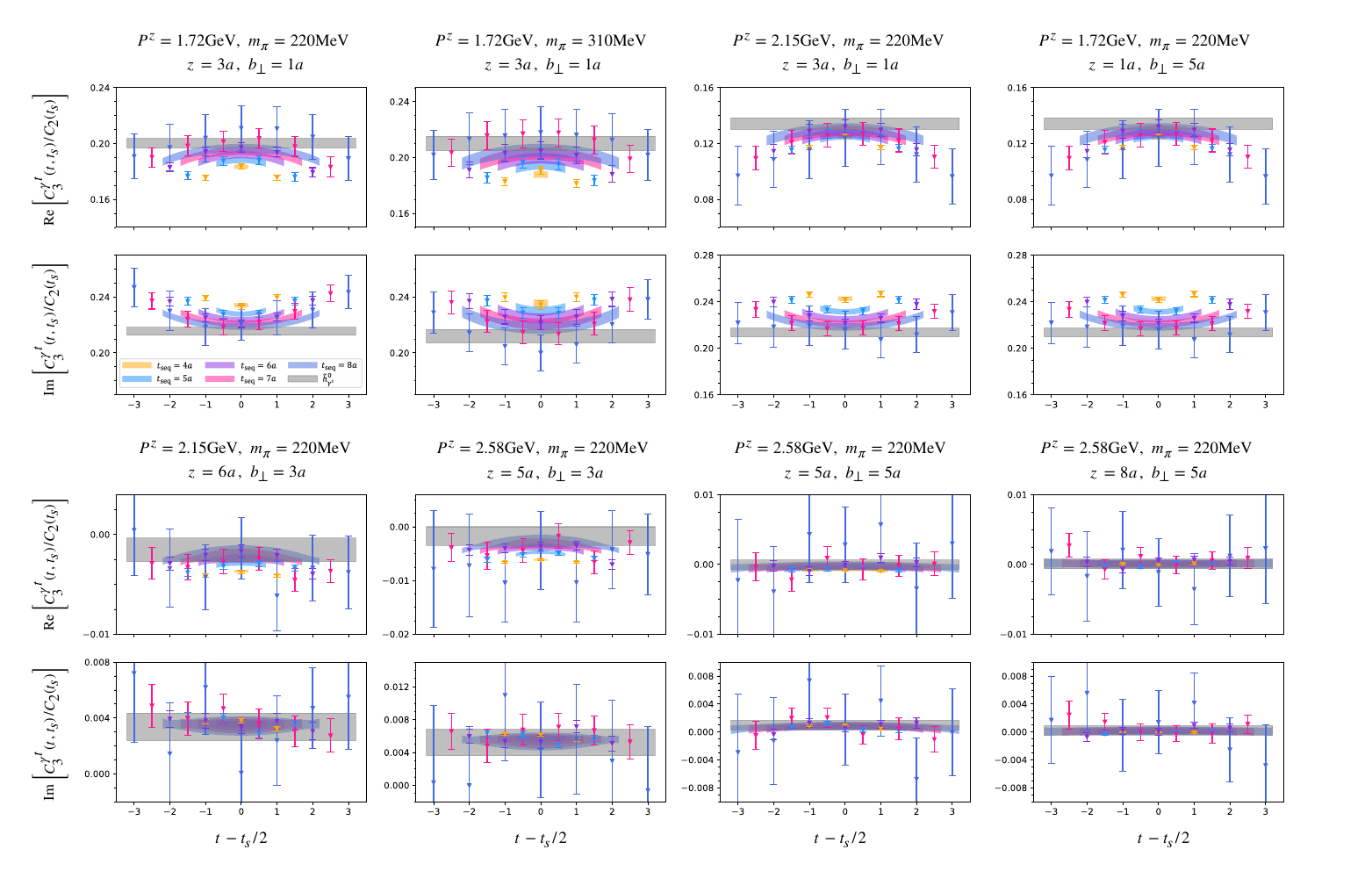}
\caption{Ratios of $C_{3}^{\gamma^t}(t, t_s)/C_2\left(t_s\right)$ (data points),  as functions of $t$ and $t_s$,  with various combination of  $\{P^z,m_{\pi}\}$ and  different $z$ and $b_{\perp}$. In this figure, the colored bands correspond to the fitted results, and the grey bands are the ground-state contribution.   }
\label{fig:Fig_JointFitRatio}
\end{figure*}
 
 \begin{figure*}[!t]
\centering
\includegraphics[scale=0.6]{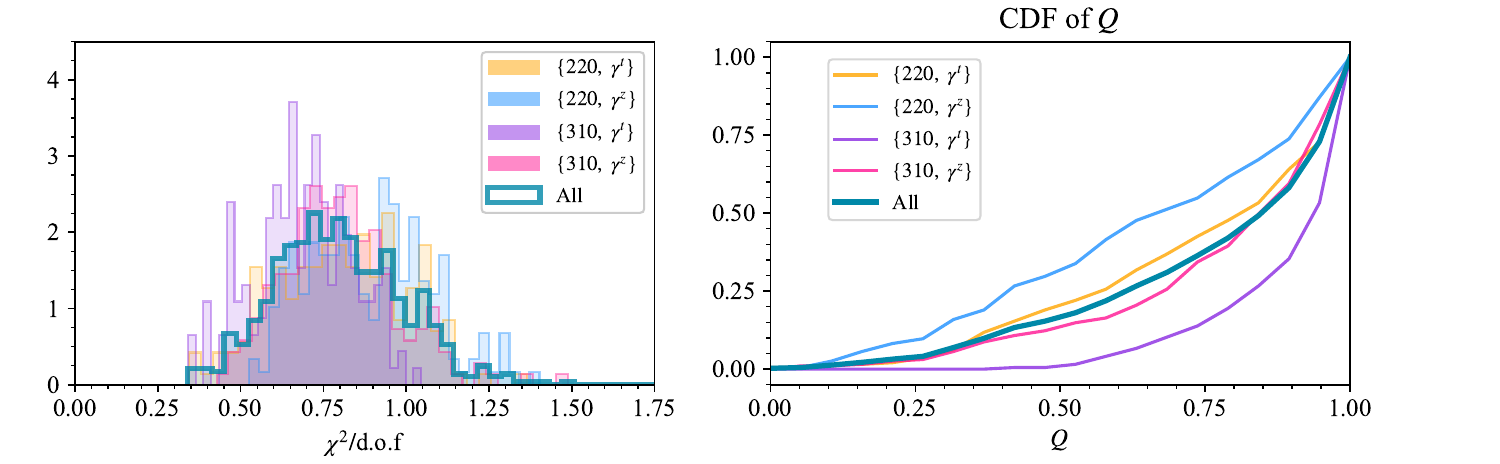}
\caption{Left panel: the $\chi^2$/d.o.f. distribution of all ground-state fits for different  combinations of $\{m_{\pi} (\mathrm{MeV}),\Gamma\}$. Right panel: the cumulative distribution function (CDF) of the Q values.  }
\label{fig:Fig_chi2_Q_distribution}
\end{figure*}

\begin{figure*}[!t]
    \centering
    \includegraphics[width=0.4\textwidth]{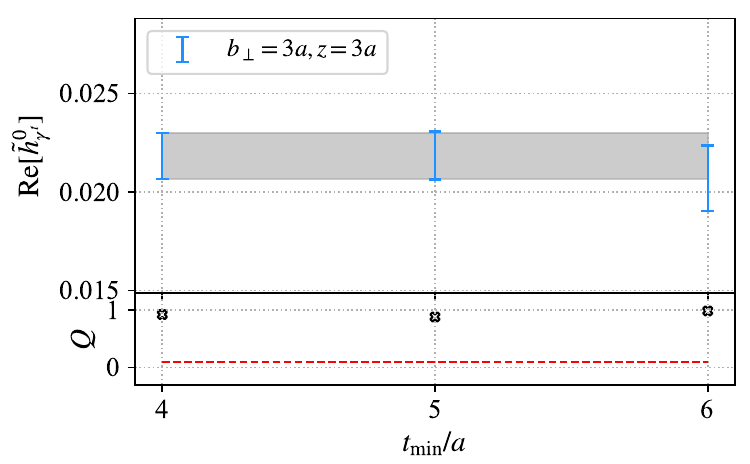}
    \quad
    \includegraphics[width=0.4\textwidth]{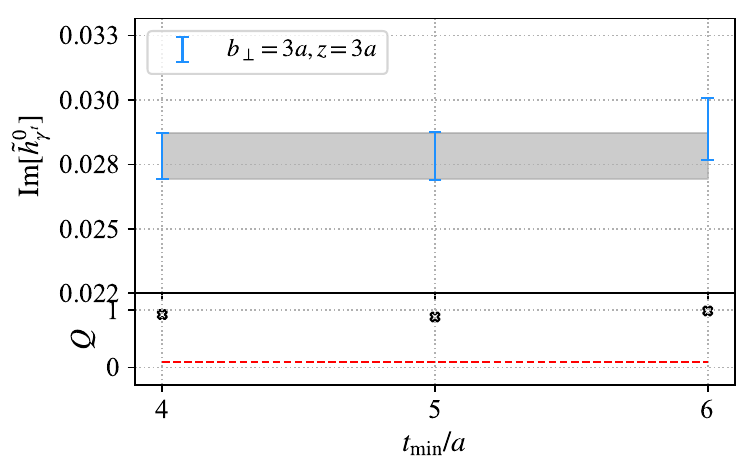}
        
    \includegraphics[width=0.4\textwidth]{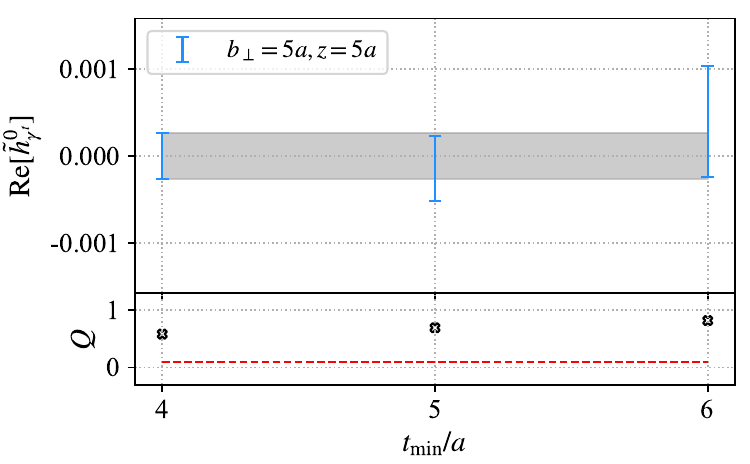}
    \quad
    \includegraphics[width=0.4\textwidth]{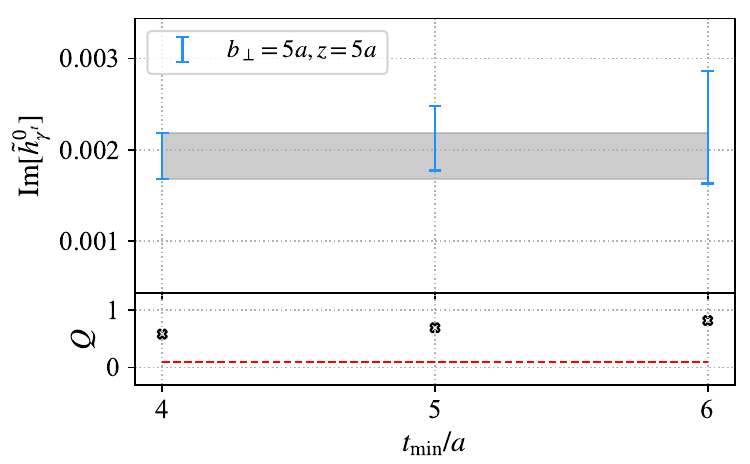}
    \caption{Stability plots to compare fit results of bare matrix elements when varying the minimum $t_s$ in the fits of $C_{3}^{\gamma^t}(t, t_s)/C_2\left(t_s\right)$. The two upper subdiagrams denote the real and imaginary results at $(b_\perp, z) = (3, 3)a$ and the lower ones are at $(5, 5)a$. The $Q$ in each subplots represents the p-value of the fits, all the fits above have good qualities with p-values greater than 0.1, which are denoted with the red dashed lines. }
    \label{fig:stability_plots}
\end{figure*}

To extract the quasi TMDPDFs, one constructs the three-point functions
\begin{align}
 &C_{3}^{\Gamma}\left(t, t_s\right) = \Big\langle \sum_{\vec{y}} e^{i\vec{P}\cdot\vec{y}} T \chi(\vec{y},t_s) \sum_{\vec{x}}  \tilde{O}_{\mathrm{TMD}}^{\Gamma}\left(\vec{x}, t\right) \bar{\chi} (\vec{0},0)     \Big\rangle.
\end{align}
We adopt the sequential source method \cite{Kilcup:1985fq} to reduce the number of propagators in three-point functions, $t_s$ denotes the time position of the sequential source. The operator $\tilde{O}_{\mathrm{TMD}}^{\Gamma}\left(\vec{x}, t\right)$ is short for the TMD nonlocal quark bilinear operator $\tilde{O}^{0}_{\sqsubset}(\vec{x}+z\hat{n}_z, \vec{x}+b_{\perp}\hat{n}_{\perp},\Gamma,  L)$ at discrete time slice $t\in[0,t_s]$. The three-momentum is chosen as $\vec{P}=(0,0, P^z)$. 

After inserting the single particle intermediate states, one can parametrize the ratio of three- and two-point functions as 
\begin{align}
    &\frac{C_{3}^{\Gamma}\left(t, t_s\right)}{C_2(t_s)} = \frac{\tilde{h}^0_{\Gamma}+c_2\big( e^{-\Delta Et} + e^{-\Delta E( t_s - t)} \big) +c_3e^{-\Delta Et_s}}{1+c_1e^{-\Delta Et_s}},
\label{eq:ratio_of_2pt_and_3pt}
\end{align}
in which $\tilde{h}^0_{\Gamma}\equiv\tilde{h}^0\left(z, b_{\perp}, P^z, \Gamma\right)$,  $\Delta E$ is the mass gap between the ground-state and excited state, and $c_{1, 2, 3}$ are parameters for the excited-state contaminations. Combining the parametrization form of 2pt functions: $C_2(t)=c_0e^{- E_0t}\left(1+c_1e^{-\Delta Et} \right)$, one can extract the values of $\tilde{h}^0_{\Gamma}$ at fixed $(z, b_{\perp}, P^z)$ through a correlated joint fit.

In our calculation, we use bootstrap resampling to establish correlations among all datasets, and the correlations are maintained consistently throughout the entire analysis. For the three-point functions, we have excluded the contact points ($t=0$ and $t=t_{\text{seq}}$) for the five values of the source-sink separation time range $t_{\text{seq}}\in (0.48 \sim 0.84)$~fm. For small $b_\perp$ ($b_\perp\leq3a$), we have further removed two points near the source and sink($t=1$ and $t=t_{\text{seq}}-1$) . 

In Fig.\ref{fig:Fig_JointFitRatio}, we give the lattice data and the fit results of the real and imaginary parts of  $C_{3}^{\gamma^t}(t, t_s)/C_2\left(t_s\right)$  with different $\{P^z,m_{\pi}\}$,  $z$ and $b_{\perp}$ as examples. As shown in the figures, the fit results (colored bands) reproduce the original lattice data  points   at each $t_s$, and the grey band corresponds to the extracted ground-state matrix element.

The fitting qualities of the bootstrap samples are illustrated in Fig.\ref{fig:Fig_chi2_Q_distribution}. The left panel shows the histogram distributions (normalized to 1) of $\chi^2$/d.o.f., and the right panel shows the cumulative distribution function (CDF) of the Q values for all correlated joint fits. 
Most of the $\chi^2$/d.o.f. spread out between $0.5$ and $1.2$, and the $Q$ value is larger than 0.05 for most fits, which  indicates that the ground-state fits are reasonable.

To demonstrate the stability of ground state fits, the comparisons between fit results of the bare matrix elements $\tilde{h}^{0}_{\gamma^t}$ with different $t_{\text{min}}$ are shown in Fig.\ref{fig:stability_plots}. The $t_{\text{min}}$ represents the minimum $t_s$ in the fits of the three-point function, and the $Q$ in the figure is the p-value of the joint fits. The two upper subdiagrams are at $b_\perp = 3a$ and $z = 3a$, for the real (left) and imaginary (right) part, respectively. The two lower panels are at $b_\perp = 5a$ and $z = 5a$. It shows that when $t_{\text{min}}$ increases, that is, when fitting with fewer data points, the fit results get larger uncertainties but remain consistent within the error range.

\subsection{Renormalization}

\begin{figure}[!t]
\centering
\includegraphics[scale=0.75]{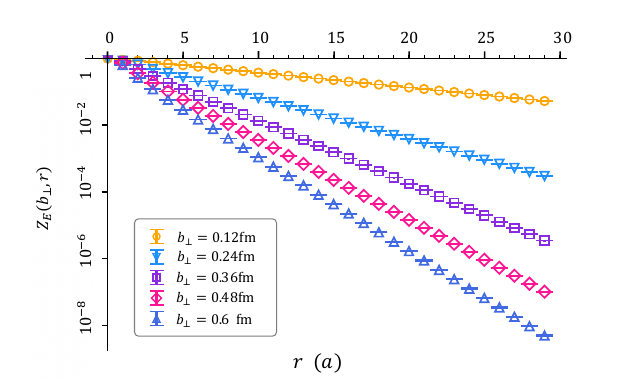}
\includegraphics[scale=0.75]{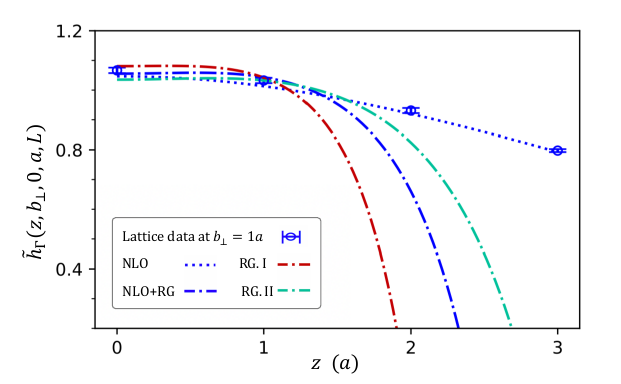}
\caption{Upper panel: Wilson loop with $r=2L+z$. 
Lower panel: renormalized zero-momentum matrix element $\tilde{h}_{\Gamma}\left(z, b_{\perp}, 0, a, L\right)$ at $b_{\perp}=1a$, compare with 1-loop results before (NLO) and after (NLO+RG) RG evolution to $\mu=2$GeV in the $\overline{\mathrm{MS}}$ scheme. Besides, we vary the starting point of evolution from  $0.8\mu_0$ (``RG. I'') to $1.2\mu_0$ (``RG. II'') to estimate the higher-order correction effects. }
\label{fig:Fig_WilsonLoop_and_ZO}
\end{figure}

As mentioned in Sec.\ref{sec:theo_renormalization}, we use the square root of the Wilson loop $\sqrt{Z_E}$ and logarithmic divergence factor $Z_O$ to renormalize the bare quasi-TMD matrix elements. In practice, the signal to noise ratio of  $Z_E(r,b_{\perp},a)$ defined in Eq.(\ref{eq:Wilson_Loop}) decreases fast with $r$ and $b_{\perp}$ such that it is hardly available for large $r$ or $b_{\perp}$. To address this, we fit the effective energies of the Wilson loop, which gives the QCD static potentials, and then extrapolate them to large $r$ and/or $b_{\perp}$, as in Ref.~\cite{Zhang:2022xuw}. Numerical results for the Wilson loop are shown in the upper panel of Fig.~\ref{fig:Fig_WilsonLoop_and_ZO}.

The logarithmic divergence factor $Z_O$ can be extracted from the matching between lattice and perturbative results in the region where both two theories work well. To preserve a good convergence of the perturbation theory before and after RG evolution, we choose the region where $b_\perp = 1a$, and $z = 0\ \text{or}\ 1a$, where both perturbation theory and lattice calculations work. The extracted $Z_O$ values at these two points are $1.0734 (93)$ and $1.0509 (92)$, respectively. Averaging these data points yields an aggregated result of $Z_O = 1.0622(87)$.

With  Eq.(\ref{eq:zo_definition}), it can be concluded that after dividing the bare matrix elements $\tilde{h}_{\Gamma}^{0}$ by $\sqrt{Z_E}$ and $Z_O$, the renormalized matrix elements   should approximately be equal to the RG evolved perturbation results $\tilde{h}_{\Gamma}^{\mathrm{\overline{MS}}}$. The lower panel of Fig.\ref{fig:Fig_WilsonLoop_and_ZO} shows the consistency at points where we extract the $Z_O$ factor, which serves as a check of the numerical result for $Z_O$. { The points at $z=0$ and $z=1$ depicted in Fig.~\ref{fig:Fig_WilsonLoop_and_ZO} demonstrate that, following renormalization with the extracted factor $Z_O = 1.0622(87)$, the renormalized matrix elements on the lattice show a consistent trend with the perturbative results obtained in $\overline{\text{MS}}$ scheme at short distance. }
To verify the accuracy of scale evolution from the intrinsic physical scale in lattice calculation to $\overline{\text{MS}}$ scale $\mu=2$GeV, we vary the origin of the evolution from $0.8\mu_0$ (shown as ``RG. I'') to $1.2\mu_0$ (shown as ``RG. II''), which is sensitive to  higher-order corrections. 
One can see that the perturbative results from different $\mu_0$ are consistent with the lattice data in the matching region ($b_\perp = 1a$, $z = 0\ \text{or}\ 1a$).

\subsection{$L$-dependence of subtracted quasi TMD matrix elements}

For a well-defined quasi TMDPDF, the length of the Wilson link $L$ should be large enough to ensure that the final results are independent of $L$.  
In the definition of the staple-shaped Wilson link in Eq.(\ref{eq:stapleshapedWilsonLink}), $L$ should be extended to infinity. The $L$-dependence is included in the linear divergence of the Wilson link self-energy and pinch-pole singularity, which can be subtracted by the square root of the Wilson loop. Therefore, the subtracted quasi TMD matrix elements
\begin{align}
	\tilde{h}_{\Gamma}(z,b_{\perp},a,L)\equiv \frac{\tilde{h}^{0}_{\Gamma}\left(z, b_{\perp}, P^z, a, L\right)}{\sqrt{Z_E\left(2L+z, b_{\perp}, a\right)} Z_O\left(1/a, {\mu}, \Gamma\right)}
\end{align}
will saturate to a plateau at large enough $L$ to ensure that the link can extend outside the region of the parton, and exhibit independence of $L$.

\begin{figure*}[!t]
\centering
\includegraphics[scale=0.7]{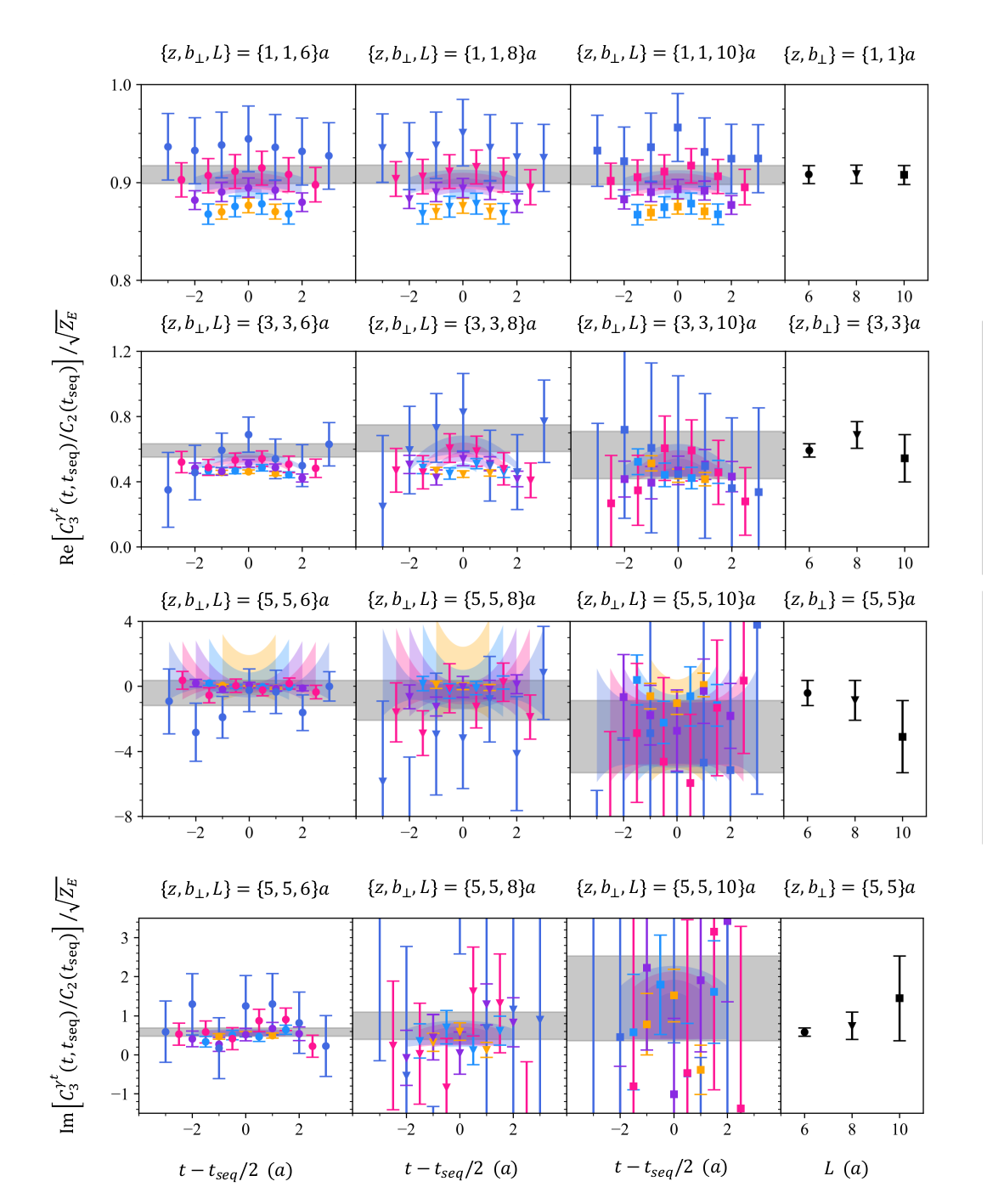}
\caption{{The real parts of ratios defined in Eq.(\ref{eq:ratio_of_2pt_and_3pt})  divided by the Wilson loop (three left panels) and the $L$ dependence of the ground state results after fitting (right panels) at $\{z,b_{\perp}\}=\{1,1\}, \{3,3\}$ and $\{5,5\}a$, with $P^z=1.72$ GeV, $m_{\pi}=220$~MeV and $\Gamma=\gamma^t$. Since the real parts for $\{z,b_{\perp}\}=\{5,5\}a$ are close to zero, we also give the imaginary parts in this example.
The labels are similar to the corresponding ones in Fig.\ref{fig:Fig_JointFitRatio}. }}
\label{fig:Fig_Ldependence}
\end{figure*}

\begin{figure*}[!th]
\centering
\includegraphics[scale=0.8]{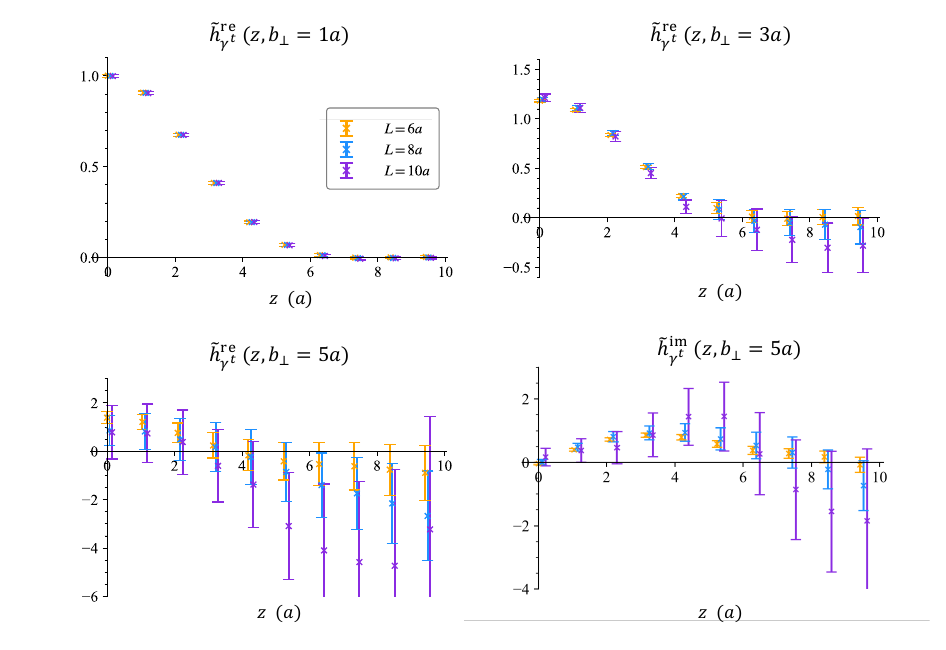}
\caption{{$L$-dependence of subtracted quasi TMDPDF matrix elements with $b_{\perp}=\{1,3, 5\}a$ at $P^z=1.72$GeV, $m_{\pi}=220$~MeV, $\Gamma=\gamma^t$. The real parts of $b_{\perp}=\{1,3\}a$ and both the real and imaginary parts of $b_{\perp}=5a$ are exhibited.}}
\label{fig:CompareLdependence_zlist}
\end{figure*}

Ref.~\cite{Zhang:2022xuw} explored the $L$-dependence of renormalized TMD matrix elements and reached the conclusion that setting $L=6a$ for MILC12 is sufficient, aligning well with the ensemble we have employed. To verify this, we further examine the $L$-dependence of the subtracted quasi TMD matrix elements, illustrated in Fig.~\ref{fig:Fig_Ldependence} and Fig.~\ref{fig:CompareLdependence_zlist}.

In Fig.\ref{fig:Fig_Ldependence}, the three panels on the left show the real parts of the ratios defined in Eq.(\ref{eq:ratio_of_2pt_and_3pt})  divided by the square root of the Wilson loop with various combinations of $z$ and $b_{\perp}$ and $L=\{6, 8, 10\}a$. The right panel gives the fit results of subtracted quasi TMD matrix elements with different $L$. From the comparison, one can see that the fitted results for $L=\{6, 8, 10\}a$ are consistent with each other. However, it should be noted that for large spatial separations, the signal becomes worse with increasing $L$. In Fig.\ref{fig:CompareLdependence_zlist}, a comparison of the subtracted quasi TMDPDF as a function of $\lambda$ with different $L$, also confirms this behavior.  Therefore, to balance the statistical and systematic uncertainties, we adopt $L=6a$ as an optimal choice in our calculation. 
 
There is a phenomenological explanation for the earlier saturation of the Wilson line; however, it is essential to note that a definitive proof is currently unavailable. When calculating the TMDPDF of a proton, which reflects the quark correlation within the proton, a reasonable estimate for the correlation length could be the size of a proton, approximately 1 fm. Beyond this distance, quarks and gluons may escape the proton, potentially diminishing their impact. This suggests a typical saturation length of around 1 fm, with $L=6a$ (or more precisely, $L=8a$) being in close proximity to this length scale.

\subsection{Quasi TMDPDFs in the coordinate space and $\lambda$ extrapolation} \label{sec:lambdaextrapolation}

\begin{figure*}[!th]
\centering
\includegraphics[scale=0.58]{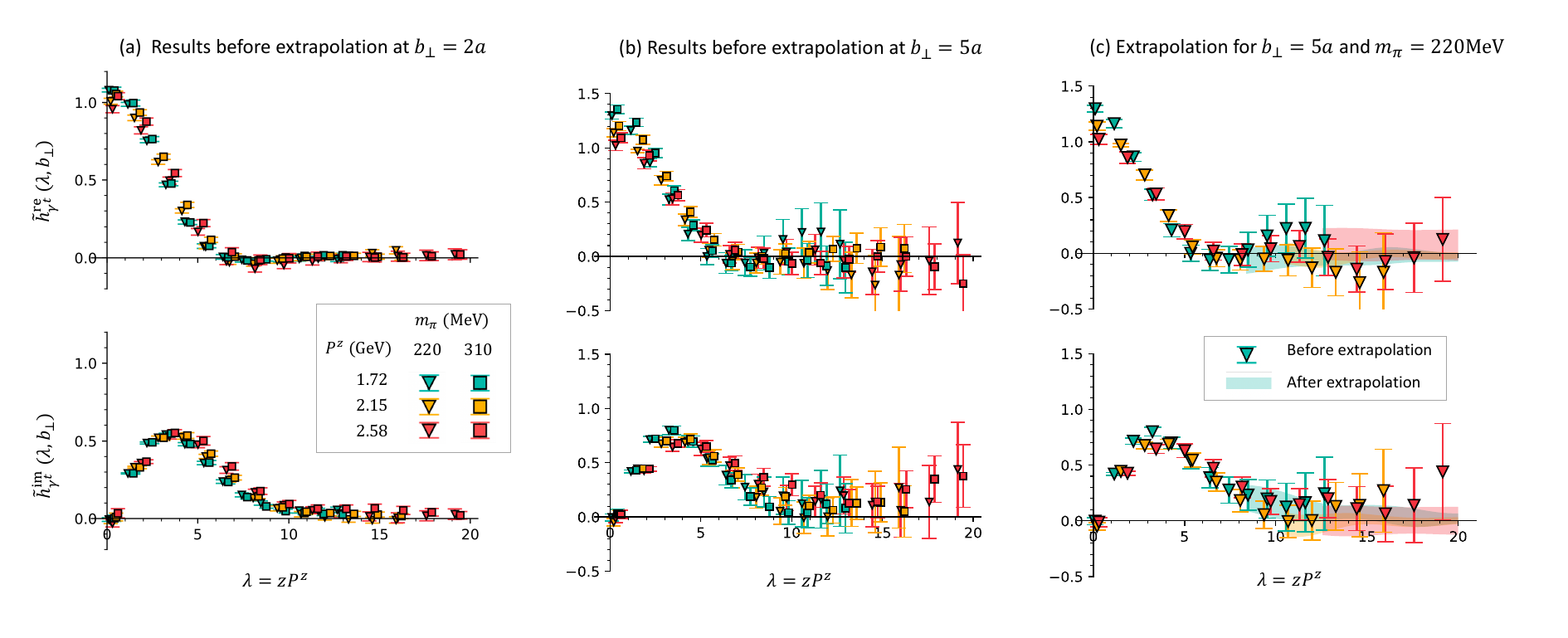}
\caption{(a, b): Renormalized quasi TMDPDFs in coordinate space as function of $\lambda=zP^z$ with various $m_{\pi}$ and $P^z$ at $b_{\perp}=2a$ and $5a$; (c): Comparison of original (data points) and extrapolated results (colored bands) at $b_{\perp}=5a$. }
\label{fig:Fig_quasiMEbeforeExtra}
\end{figure*}

\begin{table*}
\begin{center}
\renewcommand{\arraystretch}{1.5}
\setlength{\tabcolsep}{1.8mm}
\begin{tabular}{c|ccccc|c}
  \hline
$b_{\perp}$ $(a)$ & $1$ & $2$ & $3$ & $4$ & $5$ & Joint \\\hline
$n_1$         & $0.909(39)$ & $0.943 (61)$ & $0.89 (10)$ & $0.801 (78)$ & $0.84 (16)$ & $0.887 (28)$    \\
$n_2$         & $1.31 (34)$ &  $2.37 (68)$ & $1.71 (31)$ & $1.55 (38)$ & $1.22 (44)$  & $1.65 (12)$    \\
$\lambda_0$  &  $2.63 (38)$ & $3.20 (80)$ &  $2.42 (85)$  & $4.3 (1.6)$  & $4.4 (2.8)$  & $2.53 (28)$     \\\hline
$\chi^2$/d.o.f. & $1.0$ & $1.1$ & $1.3$ & $0.75$ & $0.57$ & $1.2$ \\\hline
\end{tabular}
\caption{Results for the parameters $n_1$, $n_2$ and $\lambda$ in Eq.(\ref{eq:lambdaextrapolation}) from separate fits and a joint fit, and the $\chi^2$/d.o.f. of each fit, taking the case of $m_{\pi}=220$~MeV, $\Gamma=\gamma^t$ and $P^z=1.72$GeV as an example. The results of different fit methods are consistent with each other.  }\label{tab:fitlambda}
\end{center}
\end{table*} 

\begin{figure*}[!th]
\centering
\includegraphics[scale=0.95]{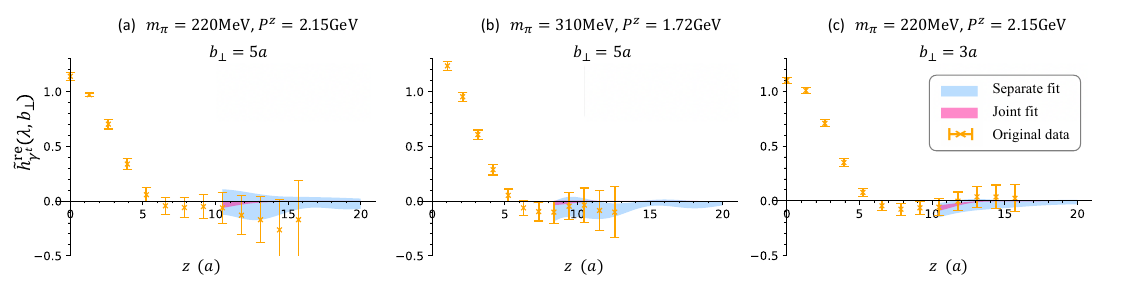}
\caption{The comparison of extrapolated results from separate fits (pale blue bands) and the joint fit (pink bands) with $\{m_{\pi}, P^z, b_{\perp}\}=\{220~\mathrm{MeV}, 2.15\mathrm{GeV}, 5a\}$ (left panel), $\{310~\mathrm{MeV}, 1.72~\mathrm{GeV}, 5a\}$ (central panel) and $\{220~\mathrm{MeV}, 2.15~\mathrm{GeV}, 3a\}$ (right panel). The results of the joint fit are consistent with the separate fits, while giving stricter restrictions for large $b_{\perp}$. }
\label{fig:comparelambdaextra}
\end{figure*}

\begin{figure*}[!th]
\centering
\includegraphics[scale=0.95]{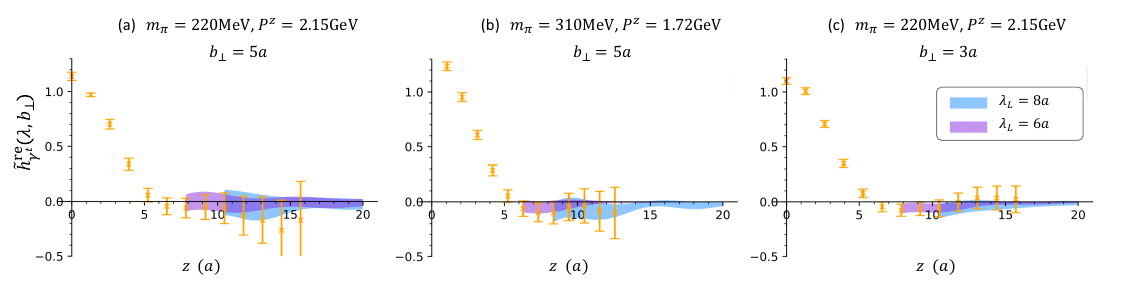}
\caption{The large $\lambda$ extrapolation of subtracted quasi TMD matrix elements in the regions $\lambda\geq\lambda_L$ with $\lambda_L=8a$ (blue bands) and $6a$ (purple bands), for the cases with $\{m_{\pi}, P^z, b_{\perp}\}=\{220~\mathrm{MeV}, 2.15~\mathrm{GeV}, 5a\}$ (left panel), $\{310~\mathrm{MeV}, 1.72~\mathrm{GeV}, 5a\}$ (central panel) and $\{220~\mathrm{MeV}, 2.15~\mathrm{GeV}, 3a\}$ (right panel). All the fitting bands agree with the original lattice data in the moderate $\lambda$ region and have smooth tails at large $\lambda$.}
\label{fig:comparelambdaextrarange}
\end{figure*}

Combining the bare quasi TMDPDFs matrix elements with the corresponding Wilson loop and renormalization factor, we obtain  numerical results for renormalized matrix elements at different $\lambda=zP^z$. 
The two panels on the left in Fig.~\ref{fig:Fig_quasiMEbeforeExtra} exhibit the $\lambda$ dependence of the renormalized matrix elements with $b_{\perp}=2a$ and $5a$ with various $P^z$ and $m_{\pi}$. 
It can be seen that as $\lambda$ increases, the quasi TMDPDFs approach zero for both real and imaginary components. When $b_{\perp}$ is large, considerable uncertainties exist in the fully correlated datasets and non-zero central values can induce unphysical oscillations for a brute-force Fourier transformation.

To address this, we adopt a physics-inspired extrapolation at large $\lambda$: \cite{Ji:2020brr}
\begin{align}
    \tilde{h}_{\Gamma, \mathrm{extra}}(\lambda) = \left[\frac{m_1}{(-i \lambda)^{n_1}}+e^{i \lambda} \frac{m_2}{(i \lambda)^{n_2}}\right] e^{-\lambda / \lambda_0}, \label{eq:lambdaextrapolation}
\end{align}
in which all parameters $m_{1,2}$, $n_{1,2}$ and $\lambda_0$ depend on the transverse separation $b_{\perp}$. The algebraic terms account for a power law behavior in the endpoint region, and the exponential term is motivated by the expectation that the correlation function has a finite correlation length (denoted as $\lambda_0$) at finite momentum.

It should be noticed that the $\lambda$ extrapolation with Eq. (\ref{eq:lambdaextrapolation}) was was initially proposed for the one-dimensional parton distribution functions~\cite{Ji:2020brr}. For TMDPDFs,  whether this form holds and whether the involved parameters depend on the transverse separation remain to be found out. In this work, we have performed the $\lambda$ extrapolation in two ways: a joint extrapolation with the same parameters and an independent extrapolation for each $b_{\perp}$.  The results are collected in Tab.\ref{tab:fitlambda}, from which one can see that each separate fit result is consistent with the joint one. However, the joint fit approach gives a more stringent constraint for the large $b_{\perp}$ case, and the corresponding result for the extrapolated matrix element has a smaller error than the original lattice data,  as shown in Fig.\ref{fig:comparelambdaextra}.  To be conservative, we have adopted the independent fits for our results.

To perform the extrapolation, a reasonable range of  $\lambda$ is required to determine the parameters. In Fig.\ref{fig:comparelambdaextrarange}, we show several extrapolations of the quasi TMD matrix elements subtracted with different combinations of $\{m_{\pi},P^z,b_{\perp}\}$. The fit is performed in the region $\lambda\geq\lambda_L$ with $\lambda_L$ being a truncation parameter.  As one can see from the figures,  the fitting bands agree with the original lattice data in the moderate $\lambda$ region and have smooth tails at large $\lambda$. With that, the quasi TMDPDFs in momentum space do not have unphysical oscillations which often show up in the brute-force Fourier transformation of $\lambda$. As a conclusion, we choose $\lambda_L=8a$ for the extrapolation and use $\lambda_L=6a$ to estimate the systematic uncertainties from $\lambda$ extrapolation. As shown in Fig.~\ref{fig:Fig_quasiMEbeforeExtra}(c), the extrapolated results (colored bands) agree with lattice data in the moderate $\lambda$ region, and give smoothly-decaying distributions at large $\lambda$.

\subsection{TMDPDFs and physical extrapolation} \label{sec:physicalextrapolation}

\begin{figure}[!th]
\centering
\includegraphics[scale=0.85]{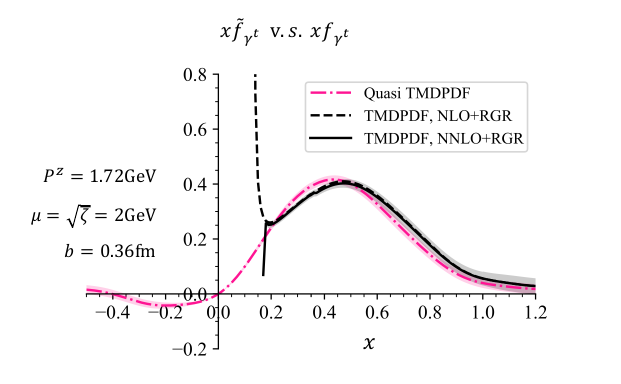}
\caption{{Quasi TMDPDF (dashed-red line) with nucleon boosted momentum $P^z=1.72$~GeV and matched TMDPDF from NLO kernel (dashed-black line, no error) and NNLO kernel (solid-black line) with RG running to scale $\mu=\sqrt{\zeta}=2$~GeV at $b_{\perp}=3a$. Only statistical errors are included in the bands. Deviations between NLO and NNLO results at $x<0.2$ indicate that the perturbative matching fails in the small-$x$ region.} }
\label{fig:Fig_compare_quasi_lc}
\end{figure}

After Fourier transforming the renormalized quasi TMDPDFs in coordinate space to momentum space, one can obtain the quasi TMDPDF $\tilde{f}_{\Gamma}(x,b_{\perp}, \zeta_z, \mu)$, shown as the  dashdotted (magenta) line in Fig.~\ref{fig:Fig_compare_quasi_lc}. Combing the updated results of Collins-Soper kernel \cite{LPC:2022ibr} and intrinsic soft function \cite{LatticePartonLPC:2023pdv} calculated on the same ensembles, we obtain the TMDPDFs through the matching formula in Eq.(\ref{eq:factoizationFormula}). Fig.~\ref{fig:Fig_compare_quasi_lc} shows an example of matched TMDPDF from the NLO \cite{Ji:2019ewn, Ji:2020ect} and NNLO \cite{delRio:2023pse, Ji:2023pba} kernel with RG running to scale $\mu=\sqrt{\zeta}=2$~GeV. One can see that the results agree within uncertainties, except for the endpoint region. 

 It should be noted that the TMDPDFs can be obtained after employing the factorization formula, while residual dependence on the lattice inputs (such as $P^z$ and $m_\pi$) may still reside in the obtained results. To  diminish  the dependence, we extrapolate the above results to the physical $m_{\pi}$ value ($135$~MeV) and  infinite momentum using the following ansatz:  
 \begin{align}
    	f(m_{\pi}, P^z)& = f_{\mathrm{phy}}\bigg[1+d_0 \left( m_{\pi}^2-m_{\pi, \mathrm{phy}}^2 \right)  \nonumber\\
     &  +d_0' \ln(m_{\pi}^2/m^2_{\pi, \mathrm{phy}})   + \frac{d_1}{\left(P^z\right)^2}+ \frac{d_1'}{P^z} \bigg], \label{eq:mpiPzextrapolation}
\end{align}
where the $d_0^{(')}$ term characterizes the pion mass dependence, and $d_1^{(')}$ accounts for the momentum-dependent discretization error.
An interesting analysis has recently derived the chiral logarithms for quasi-PDFs~\cite{Liu:2020krc}. In this approach, one first performs an operator product expansion and constructs the moments of quasi PDFs with derivative operators. Using chiral perturbation theory, one can find hadron-level operators which have the same symmetry. Calculating the one-loop perturbative contributions and summing these contributions for moment operators will lead to the chiral logarithms in quasi-PDFs. However, this analysis has not been generalized to TMDPDFs at present. In addition, since TMDPDFs are three-dimensional distributions, expanding the nonlocal operators will require the derivatives not only in the $z$ direction, but also in the transverse direction. This will likely introduce additional complexities. 

\begin{figure}[!th]
\centering
\includegraphics[scale=0.8]{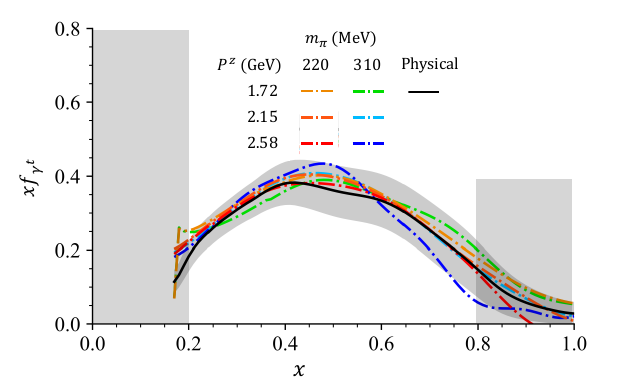}
\caption{{TMDPDFs obtained from different $m_{\pi}$ and $P^z$ (dashed-color lines) at $b_{\perp}=3a$, and their physical results after extrapolation. The shaded grey band indicates the endpoint regions where LaMET predictions are not
reliable. Only statistical errors of physical results are exhibited. }  }
\label{fig:Fig_TMDPDF_mPz_extra_both}
\end{figure}

\begin{figure}[!t]
\centering
\includegraphics[scale=0.75]{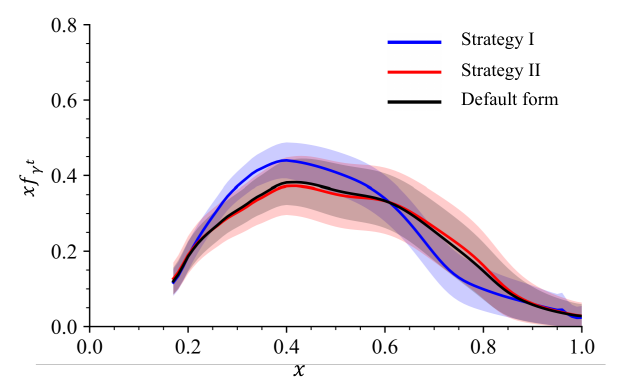}
\includegraphics[scale=0.75]{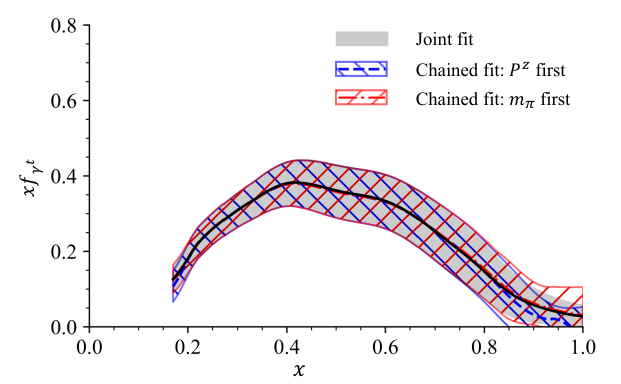}
\caption{Upper panel: Default extrapolation form in Eq.(\ref{eq:mpiPzextrapolation}), considering the systematic uncertainties from chiral extrapolation in Eq.(\ref{eq:varyingmpi}) (strategy I) and from the large-momentum extrapolation form in Eq.(\ref{eq:varyingPz}) (strategy II).
Lower panel: Comparison of the results from joint fit and chained fits with different order.}
\label{fig:Fig_chi2_chiral_Pz}
\end{figure}

Taking $d_0'=d_1'=0$ as our default case, we obtain the physical TMDPDF from a joint fit of $f(m_{\pi}, P^z)$ shown in Fig.~\ref{fig:Fig_TMDPDF_mPz_extra_both}. In order to explore the systematic bias in the extrapolation ansatz, we also estimate the $m_\pi$ dependence by using the $d_0'$ term (with $d_0=0$) and examine the $1/P^z$ contributions by adding the $d_1'$ term, shown as the following two different strategies:
\begin{itemize}
\item Default form:
\begin{align}
    	f(m_{\pi}, P^z)& = f_{\mathrm{phy}}\left[1+d_0 \left( m_{\pi}^2-m_{\pi, \mathrm{phy}}^2 \right)  + \frac{d_1}{\left(P^z\right)^2} \right], \label{eq:mpiPzextrapolation}
\end{align}
\item Strategy I: 
\begin{align}
    	&f(m_{\pi}, P^z) = f_{\mathrm{phy}}\left[1+d_0' \log\frac{m_{\pi}^2}{m_{\pi, \mathrm{phy}}^2}+ \frac{d_1}{\left(P^z\right)^2} \right], \label{eq:varyingmpi}
\end{align}
\item  Strategy II: 
\begin{align}
    	&f(m_{\pi}, P^z) = f_{\mathrm{phy}}\left[1+d_0 \left( m_{\pi}^2-m_{\pi, \mathrm{phy}}^2 \right) + \frac{d_1}{\left(P^z\right)^2} +\frac{d_1'}{P^z}  \right]. \label{eq:varyingPz}
\end{align}
\end{itemize}
Specifically, we introduce the chiral log term as well as the linear $1/P^z$ term into the fit formula, and consider the deviation between them and the default form as the systematic uncertainty.  A comparison of them is shown in the upper panel in Fig.~\ref{fig:Fig_chi2_chiral_Pz} from which one can see that the results agree with each other.

Furthermore, we have examined two different sequences of extrapolation: First, extrapolation to the physical $m_{\pi}$ followed by the $P^z\to \infty$  extrapolation, and second, the $P^z\to \infty$  extrapolation followed by the extrapolation to the physical $m_{\pi}$. A comparison of the results is presented in the lower panel of Fig.~\ref{fig:Fig_chi2_chiral_Pz}. It is evident that the results are in good agreement with each other.

\subsection{$\gamma^t-\gamma^z$ difference} \label{sec:gtgzdifference}

\begin{figure}[!th]
\centering
\includegraphics[scale=0.8]{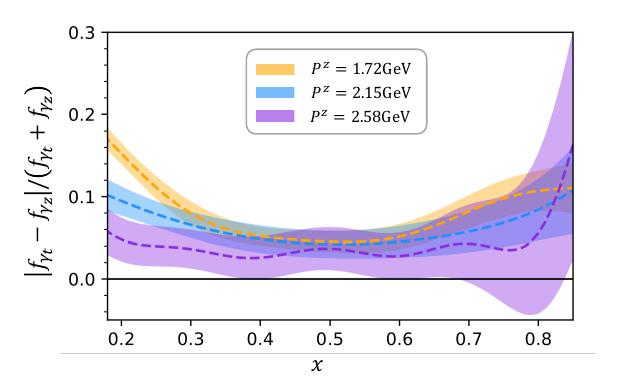}
\caption{Ratio of $|{f}_{\gamma^t}-{f}_{\gamma^z}|/({f}_{\gamma^t}+{f}_{\gamma^z})$ with $m_{\pi}=220$~MeV and different $P^z$ at $b_{\perp}=3a$
}
\label{fig:Fig_TwistCorr}
\end{figure}

In quasi TMDPDFs, both Lorentz structures $\Gamma=\gamma^t$ and $\gamma^z$ can  project onto $\gamma^+$ in the large momentum limit. 
An interesting observation  in Ref.~\cite{Constantinou:2019vyb} is that the $\gamma^t$ case has fewer operator mixing effects. 
Therefore, we use the $\Gamma=\gamma^t$ in Eq.(\ref{eq:momspacequasi}) to extract the TMDPDFs.
For the $\gamma^z$ case, in addition to the sizable operator mixing effects, it is anticipated that its deviations from the $\gamma^t$ results may come from power corrections from the operator product expansion of quasi correlators. These corrections are of order $\mathcal{O}(1/(P^z)^2)$ with opposite signs at large $P^z$.   In order to analyze the impact caused by different structures, we plot the ratio $|{f}_{\gamma^t}-{f}_{\gamma^z}|/({f}_{\gamma^t}+{f}_{\gamma^z})$ with different $P^z$ in Fig.\ref{fig:Fig_TwistCorr}.  
 One can see that with increasing nucleon momentum, the ratio becomes smaller, except for the endpoint region with  large uncertainties.  The differences will be incorporated as a systematic uncertainty.

\subsection{Estimation of systematic uncertainties}

\begin{figure}[!th]
\centering
\includegraphics[scale=0.8]{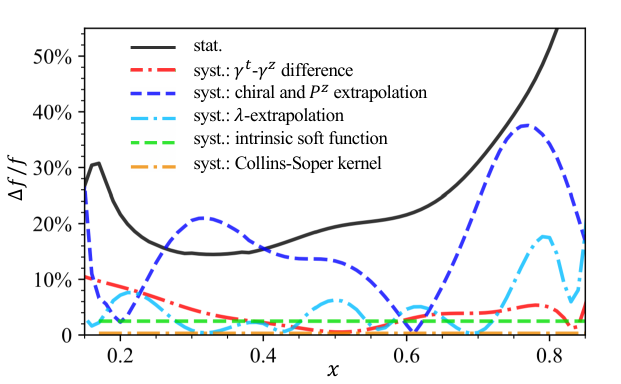}
\caption{Ratios of various uncertainties and central value of final TMDPDF at  $b_{\perp}=3a$, include the statistical one and systematical one from: (1) $\lambda$ extrapolation, (2) chiral and $P^z$ extrapolation,  (3) $\gamma^t-\gamma^z$ differences, (4) intrinsic soft function \cite{LatticePartonLPC:2023pdv} and (5) Collins-Soper kernel \cite{LPC:2022ibr}.  
}
\label{fig:Fig_AllUncertainties}
\end{figure}

As mentioned above, this work considers the systematic uncertainties from different sources, including from
\begin{itemize}
	\item different fitting ranges in the $\lambda$-extrapolation (Sec.\ref{sec:lambdaextrapolation});
	\item different strategies used for the chiral and $P^z$ extrapolation (Sec.\ref{sec:physicalextrapolation});
	\item the difference between $\gamma^t$ and $\gamma^z$ results (Sec.\ref{sec:gtgzdifference}).
\end{itemize}
In addition, we also consider the error propagating from the intrinsic soft function \cite{LatticePartonLPC:2023pdv} and Collins-Soper kernel \cite{LPC:2022ibr}, which are calculated on the same configurations. The comparison of the statistic as well as each systematic uncertainties are shown in Fig.~\ref{fig:Fig_AllUncertainties}.

\section{Final results for TMDPDFs}

\begin{figure*}[!th]
\centering
\includegraphics[scale=0.95]{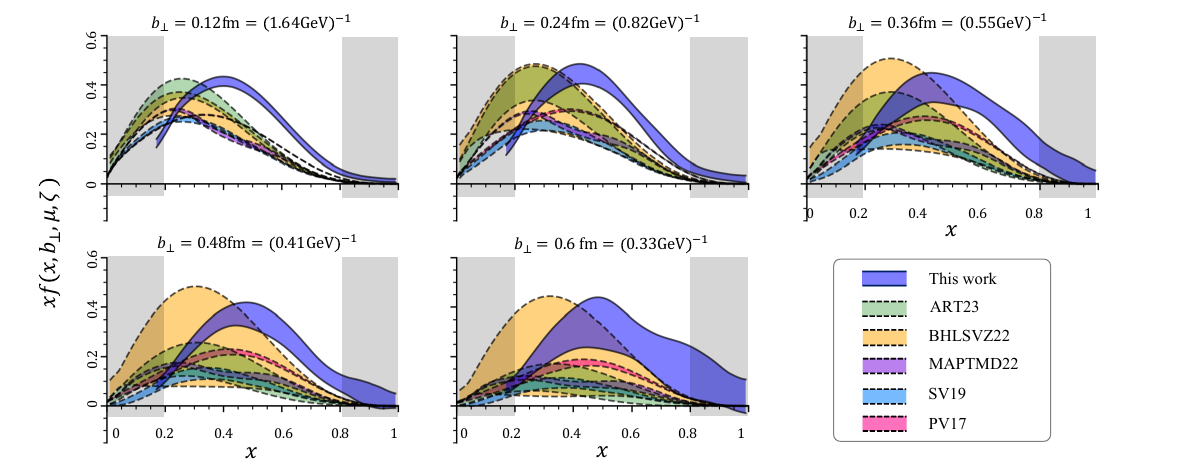}
\caption{Our final  results for unpolarized nucleon's isovector TMDPDFs $xf(x, b_{\perp}, \mu, \zeta)$ at renormalization and rapidity scales at $\mu=\sqrt{\zeta}=2~$GeV, extrapolated to physical pion mass 135~MeV and infinite momentum limit $P^z\to\infty$, compared with ART23 \cite{Moos:2023yfa}, BHLSVZ22 \cite{Bury:2022czx}, MAPTMD22 \cite{Bacchetta:2022awv}, SV19 \cite{Scimemi:2019cmh}and PV17 \cite{Bacchetta:2017gcc} global fits. 
The colored bands denote our results with both statistical and systematic uncertainties, the shaded grey regions imply the endpoint regions where LaMET predictions are not reliable. 
}
\label{fig:Fig_finalTMDPDFs}
\end{figure*}

\begin{figure*}[!th]
\centering
\includegraphics[scale=0.95]{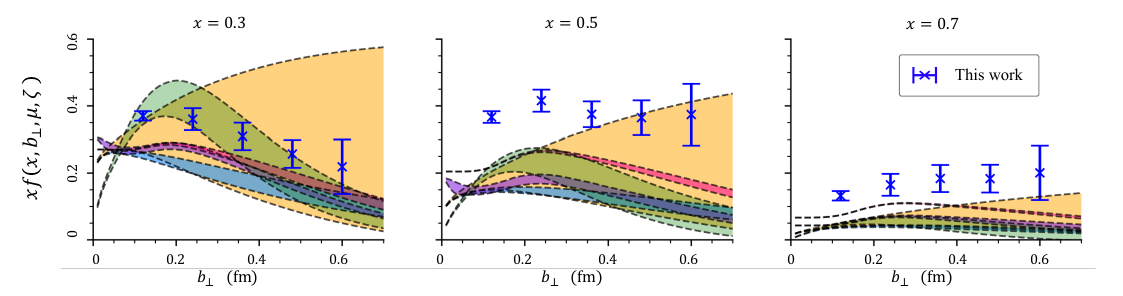}
\caption{{The TMDPDFs $xf(x, b_{\perp}, \mu=2~\mathrm{GeV}, \zeta=4~\mathrm{GeV}^2)$ with longitudinal momentum fraction $x=\{0.3, 0.5, 0.7\}$, together with the phenomenological results. The labels of the latter are the same as Fig.~\ref{fig:Fig_finalTMDPDFs}. }}
\label{fig:Fig_finalTMDPDFs_with_b}
\end{figure*}

Combing all the known uncertainties indicated in Fig.~\ref{fig:Fig_AllUncertainties}, we obtain numerical results of unpolarized nucleon's isovector TMDPDFs from our lattice simulation. Fig.~\ref{fig:Fig_finalTMDPDFs} shows $xf(x, b_{\perp}, \mu, \zeta)$ at renormalization and rapidity scales $\mu=\sqrt{\zeta}=2$~GeV as a function of $x$, together with the phenomenological results from global analyses  \cite{ Moos:2023yfa, Bury:2022czx, Bacchetta:2022awv, Scimemi:2019cmh, Bacchetta:2017gcc}.
From the comparison one can see that, our results are in qualitative agreement with  phenomenological results and share a similar behavior in $b_{\perp}$ space: the central values slowly decrease and uncertainties are gradually increasing  with the increase of $b_{\perp}$.

In Fig.~\ref{fig:Fig_finalTMDPDFs}, a bump in the $x$ distribution can be observed, representing the highest probability for the partons’ longitudinal momentum distributions. It is worth noting that the peak positions do not align precisely between the lattice results and the respective phenomenological results. Further investigations are required to elucidate this phenomenon.

The two shaded bands at the endpoint regions ($x<0.2$ and $x>0.8$) in each subplots of Fig.~\ref{fig:Fig_finalTMDPDFs} indicate that LaMET predictions are not reliable there, {which is estimated from the power correction terms $\Lambda_{\mathrm{QCD}}^2/(xP^z)^2$ and threshold logarithms $\ln((1-x)P^z)$ \cite{Braun:2018brg,Gao:2021hxl} .}  
Besides, since only one lattice spacing is used,  discretization uncertainties are not  properly handled at this stage. Especially the $b_{\perp}\sim a$ case might suffer sizable discretization effects, which can be improved by a more detailed analysis in future. 

Fig.~\ref{fig:Fig_finalTMDPDFs_with_b} shows the results for $xf(x, b_{\perp}, \mu, \zeta)$ with $x=0.3, 0.5, 0.7$ as a function of spatial separation $b_{\perp}$. The spatial distributions reflect correlations between the partons at transverse interval $b_{\perp}$ inside a highly boosted nucleon, and will reveal aspects of nucleon structure. It could be conjectured that the distributions  vanish when $b_{\perp}$  is larger than the  nucleon radius, however,  neither the present day lattice results nor the phenomenological results are precise enough to draw such a conclusion.

\section{Summary and prospect}

In summary, we have performed the first calculation of Transverse Momentum-Dependent Parton Distribution Functions (TMDPDFs) within a nucleon using the LaMET expansion of lattice data. State-of-the-art techniques in renormalization and extrapolation on the lattice have been employed, including the consideration of the perturbative kernel up to NNLO with RG resummation. We investigate the dependence on pion mass and hadron momentum, incorporating both statistical and systematic errors to provide a robust characterization of the inner structure of nucleons through the parton distributions.

While the current findings are promising, further improvements are required. Firstly, a comprehensive analysis of multiple ensembles with varying lattice spacings, pion masses, and volumes is essential. This approach will not only reassess all uncertainties addressed in this study but also systematically explore additional significant factors. Secondly, our results exhibit notable theoretical uncertainties in the endpoint regions, which could stem from power corrections or threshold logarithms. Thirdly, we expect a decay of the TMDPDF with increasing $b_{\perp}$ which is not (yet) visible in our data. Going to larger $b_{\perp}$ is thus imperative. Moreover, the presence of chiral logarithms in the analysis of quasi PDFs (as demonstrated in Ref.~\cite{Liu:2020krc}) highlights the potential for similar logarithms in our case, warranting dedicated theoretical investigation.

\section*{Acknowledgements}

We thank Alessandro Bacchetta, Matteo Cerutti, Alexey Vladimirov for sharing their data for this work.
We also thank the MILC collaboration for providing us their gauge configurations with dynamical fermions.
This work is supported in part by Natural Science Foundation of China under grant No.  12125503, 12375069, 12335003, and the Natural Science Foundation of Shandong province under the Grant No. ZR2022ZD26.  The computations in this paper were run on the Siyuan-1 cluster supported by the Center for High Performance Computing at Shanghai Jiao Tong University, and Advanced Computing East China Sub-center. The numerical calculation in this study were also carried out on the ORISE Supercomputer, and HPC Cluster of ITP-CAS. {J.H. is partially support by the U.S. Department of Energy, Office of Science, Office of Nuclear Physics under the umbrella of the Quark-Gluon Tomography (QGT) Topical Collaboration with Award DE-SC0023646.} X.J. and Y.S. are supported by the U.S. Department of Energy, Office of Science, Office of Nuclear Physics, under contract number DE-SC0020682. Y.Y. is supported in part by the Strategic Priority Research Program of Chinese Academy of Sciences, Grant No. XDB34030303 and XDPB15. J.Z. is supported in part by National Natural Science Foundation of China under grant No. 11975051. A.S, W.W, Y.Y and J.Z are also supported by a NSFC-DFG joint grant under grant No. 12061131006 and SCHA~458/22. Q.Z. is supported by the Key Laboratory of Particle Astrophysics and Cosmology, Ministry of Education of China.

\end{document}